\documentclass{aastex} 
\usepackage{emulateapj5,apjfonts,psfig,epsfig}
\begin{document} 

\slugcomment{AJ, in press}

\title{The Dwarf Spheroidal Companions to M31:  Variable Stars in 
Andromeda~II\footnotemark} 
\footnotetext{Based on observations with the NASA/ESA {\it Hubble Space 
Telescope}, obtained at the Space Telescope Science Institute, which is 
operated by the Association of Universities for Research in Astronomy, Inc., 
(AURA), under NASA Contract NAS 5-26555.}

\shortauthors{Pritzl et al.}
\shorttitle{And~II Variable Stars}

\received{}
\accepted{}
\revised{}

\author{Barton J. Pritzl and Taft E. Armandroff} 
\affil{National Optical Astronomy Observatory, P.O. Box 26732, Tucson, AZ 
85726 \\email: pritzl@noao.edu, armand@noao.edu} 
\author{George H. Jacoby} 
\affil{WIYN Observatory, P.O. Box 26732, Tucson, AZ, 85726 \\email: 
gjacoby@wiyn.org}
\author{G. S. Da Costa} 
\affil{Research School of Astronomy and Astrophysics, Institute of 
Advanced Studies, The Australian National University, Cotter Road, Weston, 
ACT 2611, Australia \\email: gdc@mso.anu.edu.au}

\begin{abstract}

We present the results of a variable star search in Andromeda~II, a dwarf 
spheroidal galaxy companion to M31, using {\it Hubble Space Telescope} 
Wide Field Planetary Camera~2 observations.  Seventy-three variables 
were found, one of which is an anomalous Cepheid while the others are
RR~Lyrae stars.  The anomalous Cepheid has properties consistent with those found 
in other dwarf spheroidal galaxies.  For the RR~Lyrae stars, the mean periods 
are 0.571~day and 0.363~day for the fundamental mode and first-overtone 
mode stars, respectively.  With this fundamental mode 
mean period and the mean metallicity determined from the red giant branch 
($\langle {\rm [Fe/H]} \rangle = -1.49$), Andromeda~II follows the 
period-metallicity relation defined by the Galactic globular clusters and 
other dwarf spheroidal galaxies.  We also find that the properties of the 
RR~Lyrae stars themselves indicate a mean abundance that is consistent with 
that determined from the red giants.  There is, however, a significant spread 
among the RR~Lyrae stars in the period-amplitude diagram, which is possibly 
related to the metallicity spread in Andromeda~II
indicated by the width of the red giant branch in Da~Costa et al.
In addition, the abundance distribution of the RR~Lyrae stars is notably wider 
than the distribution expected from the abundance determination errors alone.  
The mean magnitude of the RR~Lyrae stars, $\langle V_{\rm RR} \rangle = 
24.87\pm0.03$, implies a distance $d = 665\pm20$~kpc to Andromeda~II.\@  
This matches the distance derived from the mean magnitude of the horizontal 
branch stars by Da~Costa et al., $d = 680\pm20$~kpc. We also demonstrate 
that the specific frequency of anomalous Cepheids in dwarf spheroidal galaxies 
correlates with the mean metallicity of their parent galaxy, and that the 
Andromeda~II and Andromeda~VI anomalous Cepheids appear to follow the same 
relation as those in the Galactic dwarf spheroidals.

\end{abstract}


\section{Introduction} 

There have been many studies of the stellar populations of the Galactic dwarf 
spheroidal (dSph) galaxies (see Mateo 1998 and references therein) which 
have enabled astronomers to better understand chemical evolution and star 
formation histories in less complex environments than for spiral or 
elliptical galaxies.  In particular, the nature of dwarf galaxies has become 
important for understanding the formation of galaxies in general, as it is now 
believed that the halos of more massive galaxies were formed, at least in part, 
by the ``cannibalism'' of dwarf galaxies.  This process continues today, with 
the accretion of the Sagittarius dSph galaxy into the halo of our own 
Galaxy being a prime example (Ibata, Gilmore, \& Irwin 1994).

An important method for investigating the properties of dSph galaxies 
is to determine the parameters of their variable stars.  For example, the 
simple presence of RR~Lyrae stars (RRLs) is indicative of an older stellar 
population (age $> 10$~Gyr).  Due to their nearly uniform luminosities, RRLs 
can also be used to determine the distance to the system in which they are 
found.  Every dSph galaxy surveyed for variable stars has also been found 
to contain at least one anomalous Cepheid (AC).\@  This type of Cepheid 
variable derives its name from having a period-luminosity relationship 
that does not follow either the classical Cepheid or Population~II 
Cepheid relationships.  ACs are believed to be either young, metal-poor 
stars or stars that have formed from mass-transfer in a binary system 
(see, for example, the discussion in Pritzl et al.\ 2002, hereafter 
Paper~I).\@  

The dSph companions of the Milky Way Galaxy have all been surveyed for 
variable stars.  Recently, we have begun a variability survey of the dSph 
companions to the Andromeda Galaxy using the {\it Hubble Space Telescope} 
(HST).\@  We found the dSph galaxy Andromeda~VI (And~VI) to contain a significant 
population of RRLs and a number of ACs (Paper~I).\@  In this paper we present 
the results of our survey of Andromeda~II (And~II).\@  And~II, while having a 
mean metallicity slightly higher than that of And~VI, has been shown 
to have a substantial metallicity spread (C\^{o}t\'{e}, Oke, \& Cohen 1999; 
Da~Costa et al.\ 2000, hereafter DACS00).  Here we show that the RRLs in And~II 
also appear to have a broad metallicity distribution.  We also discuss some of 
the other properties traced out by the RRLs and present the discovery of one 
AC.\@  The relation of the AC specific frequency and the mean metallicity 
of the parent galaxy is also examined.

\begin{figure*}[t]
 \centerline{\psfig{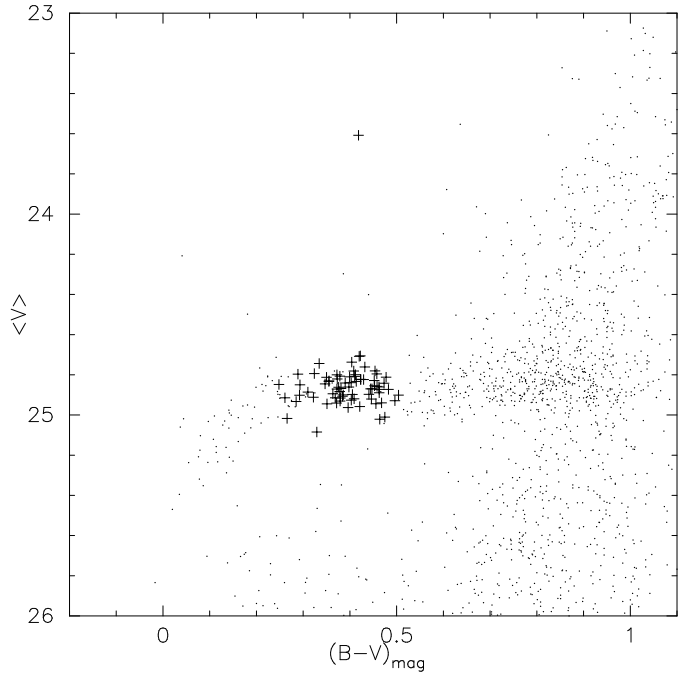}}
 \caption{And~II color-magnitude diagram.  The RR~Lyrae stars and the 
  anomalous Cepheid are shown as plus symbols.}
 \label{Fig01}
\end{figure*}

\section{Observations and Reductions} 

Observations of And~II were taken by the HST Wide Field Planetary Camera~2 
(WFPC2) as part of the GO Program 6514 on 1997 August 29 and 1997 September 3, 
using the same orientation.  Three 1200~s integrations through the F555W 
filter and seven 1300~s integrations through the F450W filter were taken during 
the first set of observations.  Identical exposure times were used for the 
second set of observations, but there were four F555W and eight F450W 
integrations.  The images were offset from the center of the dSph galaxy in 
order to avoid bright foreground stars (cf.\ Fig.~1 of DACS00). 

The point-spread function fitting photometry was performed with {\sc allframe} 
(Stetson 1994) as described in Paper~I.\@  The aperture and charge-transfer 
efficiency corrections along with the $B$,$V$ calibrations were carried out in 
the same fashion as outlined in Paper~I.\@  The main difference in the 
calibration of the And~II dataset is the availability of the photometry from 
the combined WFPC2 images by DACS00, which form the basis of their 
color-magnitude diagram (CMD).\@  We were able to compare our photometric 
calibrations with that of DACS00.  

The results of this comparison are presented in Table~1.  Here we list, for 
each wide-field CCD, the mean magnitude difference (in the sense this study -- 
DACS00) and the standard deviation of the mean of the differences, along 
with the magnitude range over which the comparison was made, and  the number 
of stars.  Candidate variables have been excluded from the calculation.  
There are obvious systematic offsets between the two sets of photometry 
but it is reassuring to note that there are no signs of any trend with 
magnitude.  The origin of these differences is unclear although, given 
the different photometry techniques and calibration processes employed 
(see Paper~I and DACS00), the existence of such zero point differences 
is not particularly surprising.  DACS00 demonstrated that 
their WFPC2 photometry agrees well with independently calibrated ground-based 
photometry for stars in the field of And~II\@.  Consequently, we have chosen 
to adjust the photometry of this study by the mean differences given in 
Table~1 to align it with that of DACS00.

\begin{figure*}[t]
  \centerline{\psfig{figure=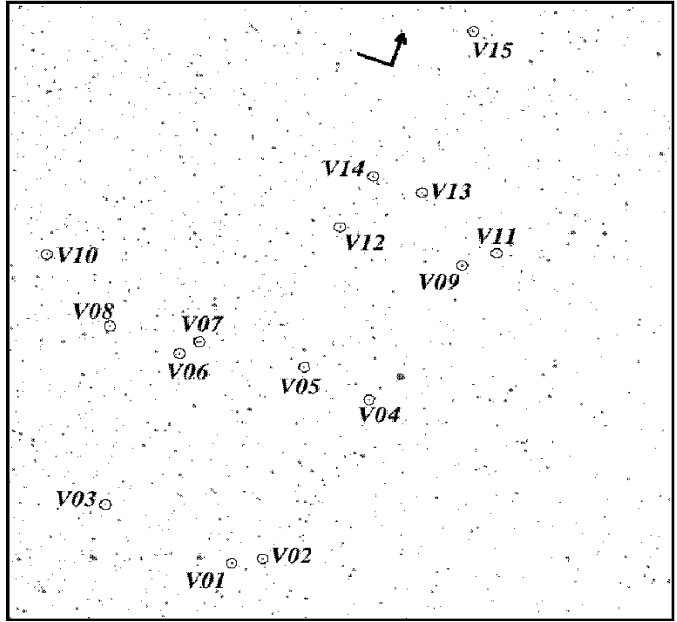,height=3.25in,width=3.50in}}
  \caption{Finding charts for the And~II variable stars.  The WFC2 
   ($1.2\arcmin$x$1.3\arcmin$), WFC3 ($1.3\arcmin$x$1.2\arcmin$), 
   and WFC4 ($1.2\arcmin$x$1.2\arcmin$) images are each shown in a 
   panel.  North and east directions are shown with the arrow pointing 
   toward the north.}
  \label{Fig02a}
\end{figure*}

\begin{figure*}[t]
  \figurenum{2 cont}
  \centerline{\psfig{figure=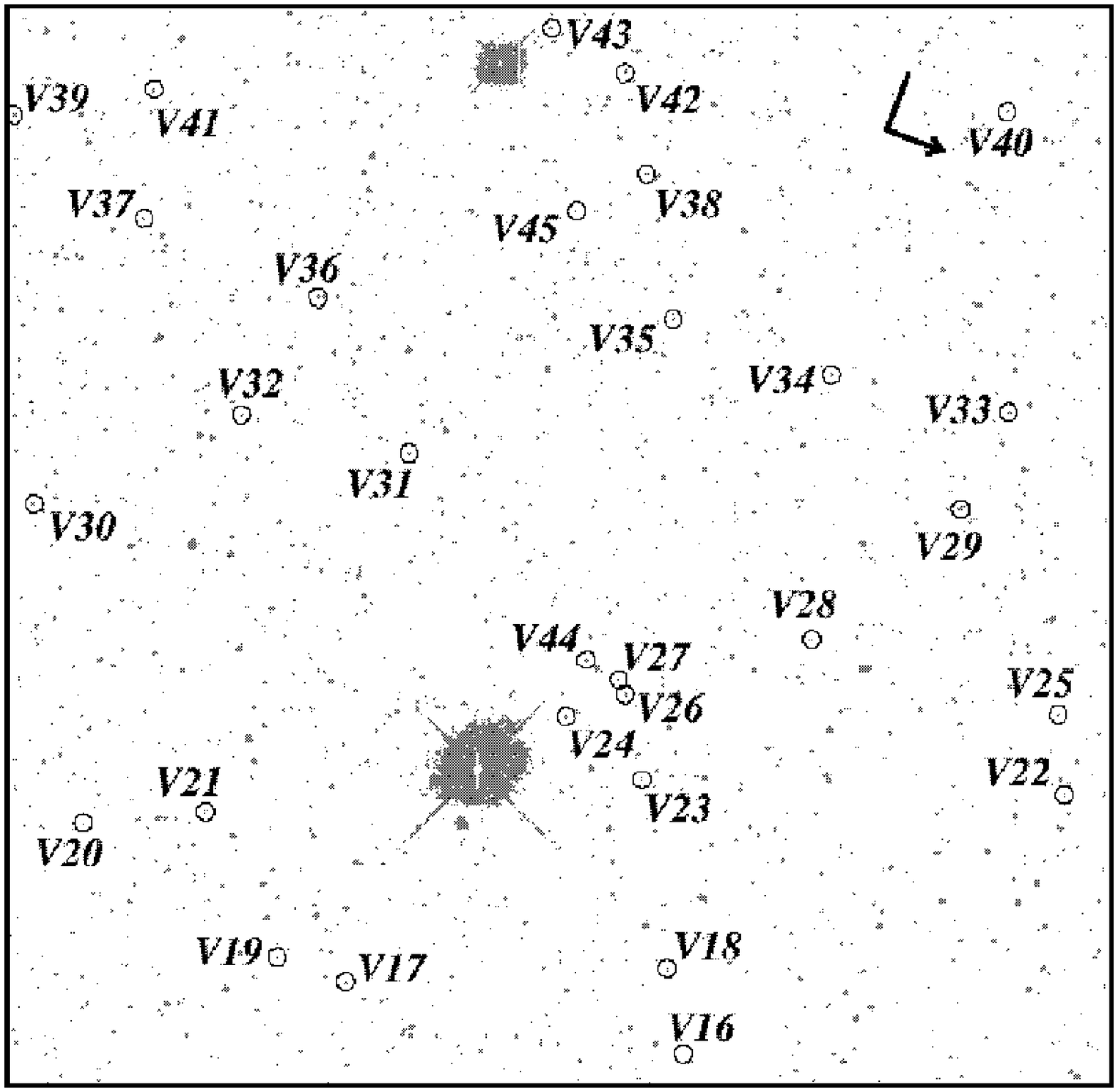,height=3.25in,width=3.50in}}
  \caption{Finding charts for the And~II variable stars.  The WFC2
   ($1.2\arcmin$x$1.3\arcmin$), WFC3 ($1.3\arcmin$x$1.2\arcmin$),
   and WFC4 ($1.2\arcmin$x$1.2\arcmin$) images are each shown in a
   panel.  North and east directions are shown with the arrow pointing
   toward the north.}
  \label{Fig02b}
\end{figure*}

\begin{figure*}[t]
  \figurenum{2 cont}
  \centerline{\psfig{figure=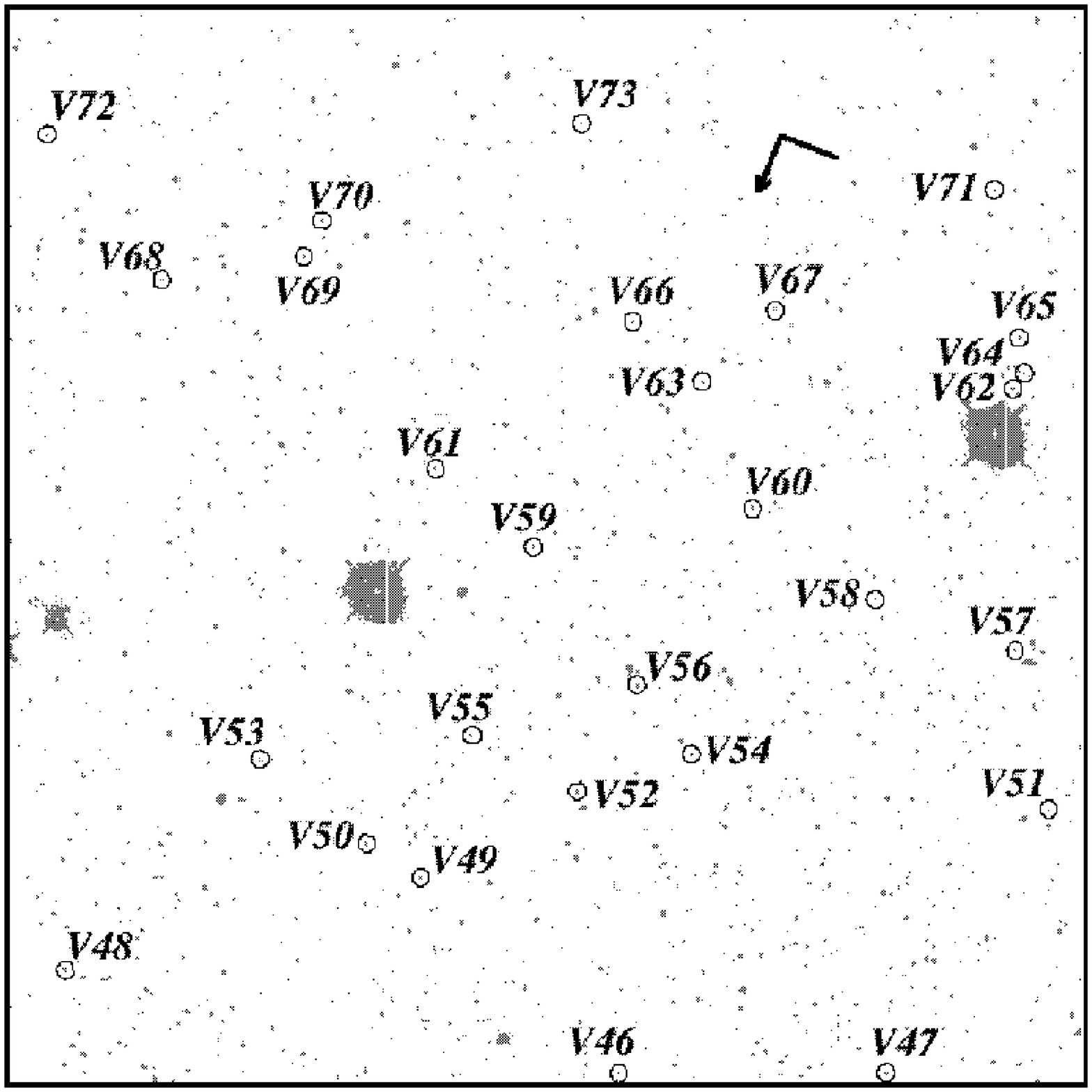,height=3.25in,width=3.50in}}
  \caption{Finding charts for the And~II variable stars.  The WFC2
   ($1.2\arcmin$x$1.3\arcmin$), WFC3 ($1.3\arcmin$x$1.2\arcmin$),
   and WFC4 ($1.2\arcmin$x$1.2\arcmin$) images are each shown in a
   panel.  North and east directions are shown with the arrow pointing
   toward the north.}
  \label{Fig02c}
\end{figure*}

\section{Variable Stars} 

As for Paper~I, we searched the photometry data for 
variable stars using a routine created by Dr.\ Peter Stetson called 
{\sc daomaster}.  This routine compares the rms scatter in the magnitudes to 
that expected from the photometric errors returned by the {\sc allframe} 
program.  There were only a handful of RRLs found on the PC and no ACs.  
Because the PC covers only a small area of sky compared to the WF CCDs, and 
because of the lack of sufficiently bright stars for reliable aperture 
corrections on the PC, we do not include the PC data in our results.

Figure~1 shows the location of the variable stars in the CMD created by 
DACS00.  We removed the DACS00 photometric data from the CMD for those 
stars which were found to be variable in this paper and instead plotted them 
as plusses representing the mean magnitude and color of the variable star.  
The RRLs can be seen filling in the gap in the CMD along the horizontal 
branch (HB).\@  About 1.2~mag brighter than the HB is the single AC we 
found in this survey.  The periods, magnitudes, and light curves were 
determined using the programs created by A. C. Layden as outlined in Paper~I.  
We note that due to the method we have used to calibrate the data to the 
$B$, $V$ system, the colors and magnitudes for the variable stars have 
inherent uncertainties on the order of 0.02-0.03~mag.  Not withstanding this 
fact, we find that our method allows us to analyze the general properties 
of the And~II variable star population.

The stars that lie to the blue of the red giant branch and 
those DACS00 stars found within and near the instability strip on the HB 
in Figure~1 were examined for variability, but no additional variables were 
found.  Two plausible explanations for the non-variability, especially for 
those stars found within the instability strip, are either (a) these 
stars are not variables, but errors in the photometry, particularly in 
the colors, have scattered them into an apparent location in the 
instability strip, or (b) these are indeed variables but the limited 
time coverage of the observations meant that we did not detect the 
variability.  

DACS00 performed an initial variable star search of And~II.\@  In their paper 
they report detecting approximately 30 candidate variable stars.  Fig.~3 
of DACS00 shows the location of these variables in the And~II CMD and it 
shows a lot of dispersion among the RRLs.  As discussed in Paper~I 
the reason for such a dispersion is that the magnitudes and colors plotted 
by DACS00 come from 2 pairs of combined frames which, consequently, do not 
correctly sample the variable star light curves.  They also show 
light curves for four of the RRLs they detected (see Fig.~6 of DACS00).  
These RRLs match V57, V60, V05, and V33 in this survey, going from the 
top-down in their figure.  The periods for V57 and V05 match well in both 
studies, while the periods found for V60 and V33 in this survey and DACS00 
are aliases of each other.  The method used in this paper to obtain the 
photometry for the variable stars results in having more data points than were 
available to DACS00 allowing a more reliable choice of the period here.

A total of 73 variable stars were found on the WFC chips, one of which is 
an AC.\@  This star will be discussed in the following section.  Table~2 
presents the photometric properties for these variable stars while their 
individual photometric $B$ and $V$ data are in Tables~3 and 4.  Column~1 of 
Table~2 lists the star's ID, while the next two columns give the RA and Dec.  
Finding charts for the variable stars are shown in Figure~2.  Column~4 lists 
the period of each star.  The intensity-weighted $\langle V \rangle$ and 
$\langle B \rangle$ magnitudes along with the magnitude-weighted colors 
$(\bv)_{\rm mag}$ are shown in columns 5-7.  Columns~8 and 9 give the $V$ and 
$B$ amplitudes of the variable stars.  The remaining columns will be discussed 
later in the paper.  In Figure~3 we present the light curves for all of the 
variable stars.

\begin{figure*}[t]
  \centerline{\psfig{figure=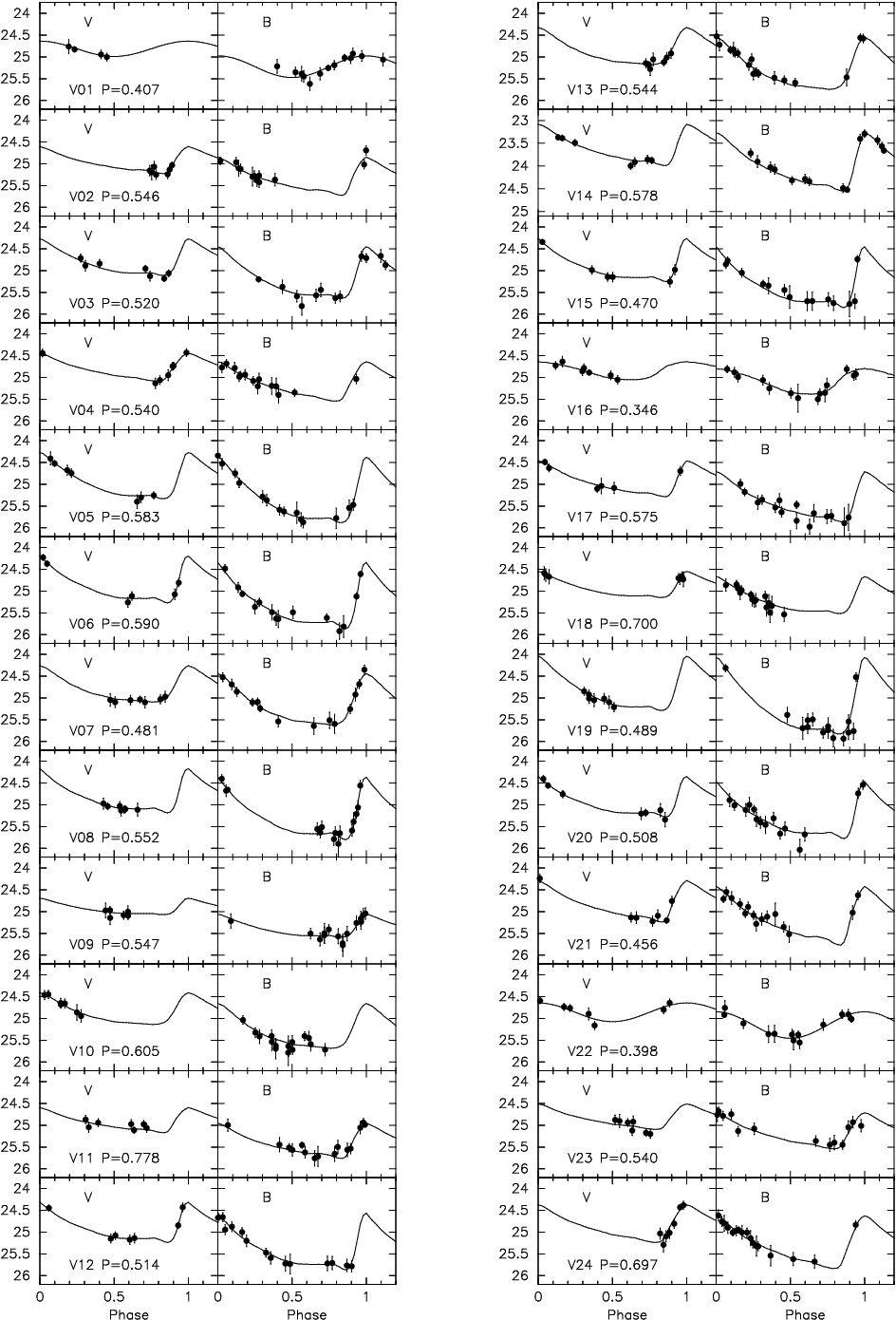,height=7.00in,width=5.00in}}
  \caption{And~II variable star light curves.  The observations are shown 
   as filled circles and the fitted templates are displayed as curves.}
  \label{Fig03}
\end{figure*}

\begin{figure*}[t]
  \figurenum{3 cont}
  \centerline{\psfig{figure=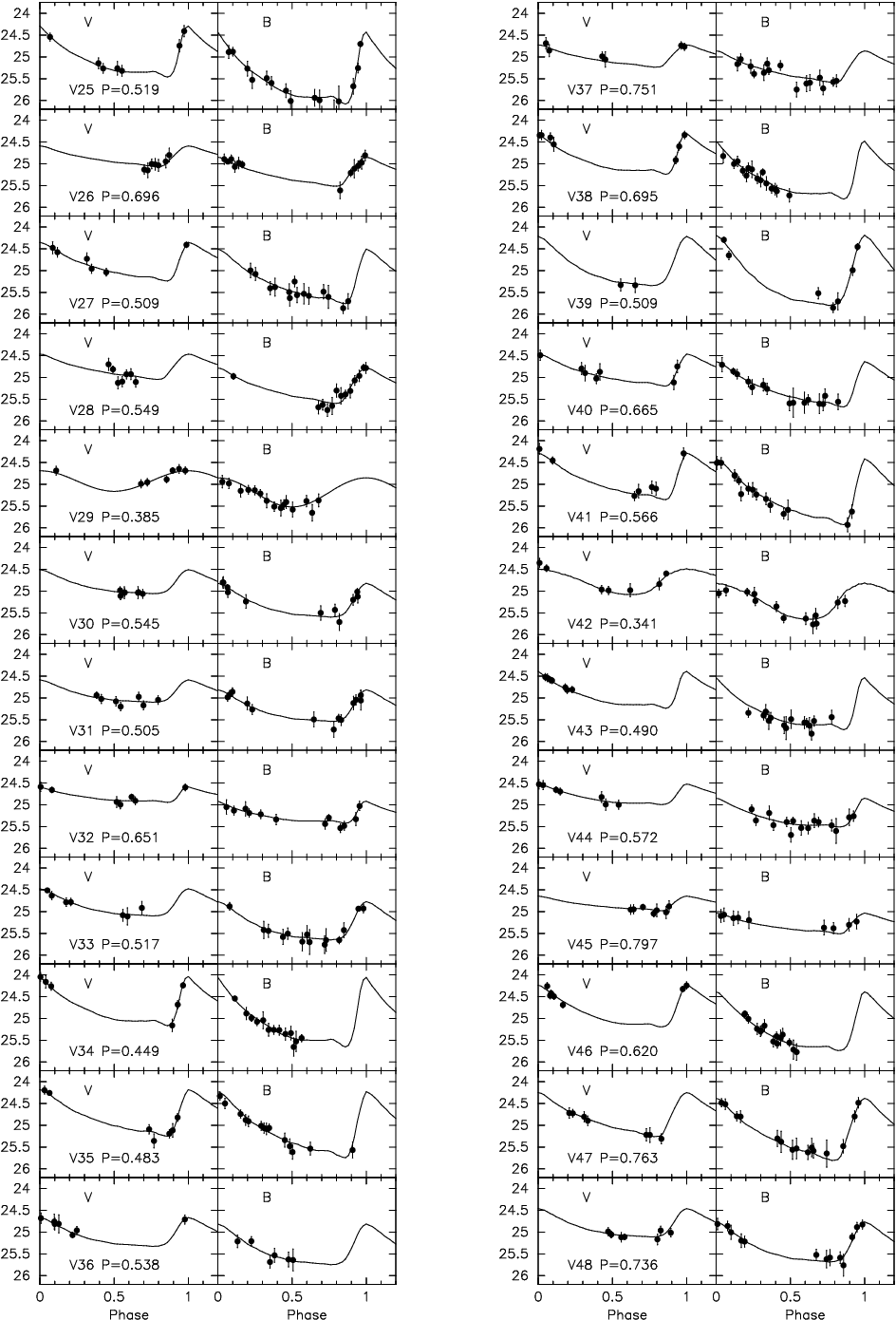,height=7.00in,width=5.00in}}
  \caption{And~II variable star light curves.  The observations are shown
   as filled circles and the fitted templates are displayed as curves.}
  \label{Fig03}
\end{figure*}

\begin{figure*}[t]
  \figurenum{3 cont}
  \centerline{\psfig{figure=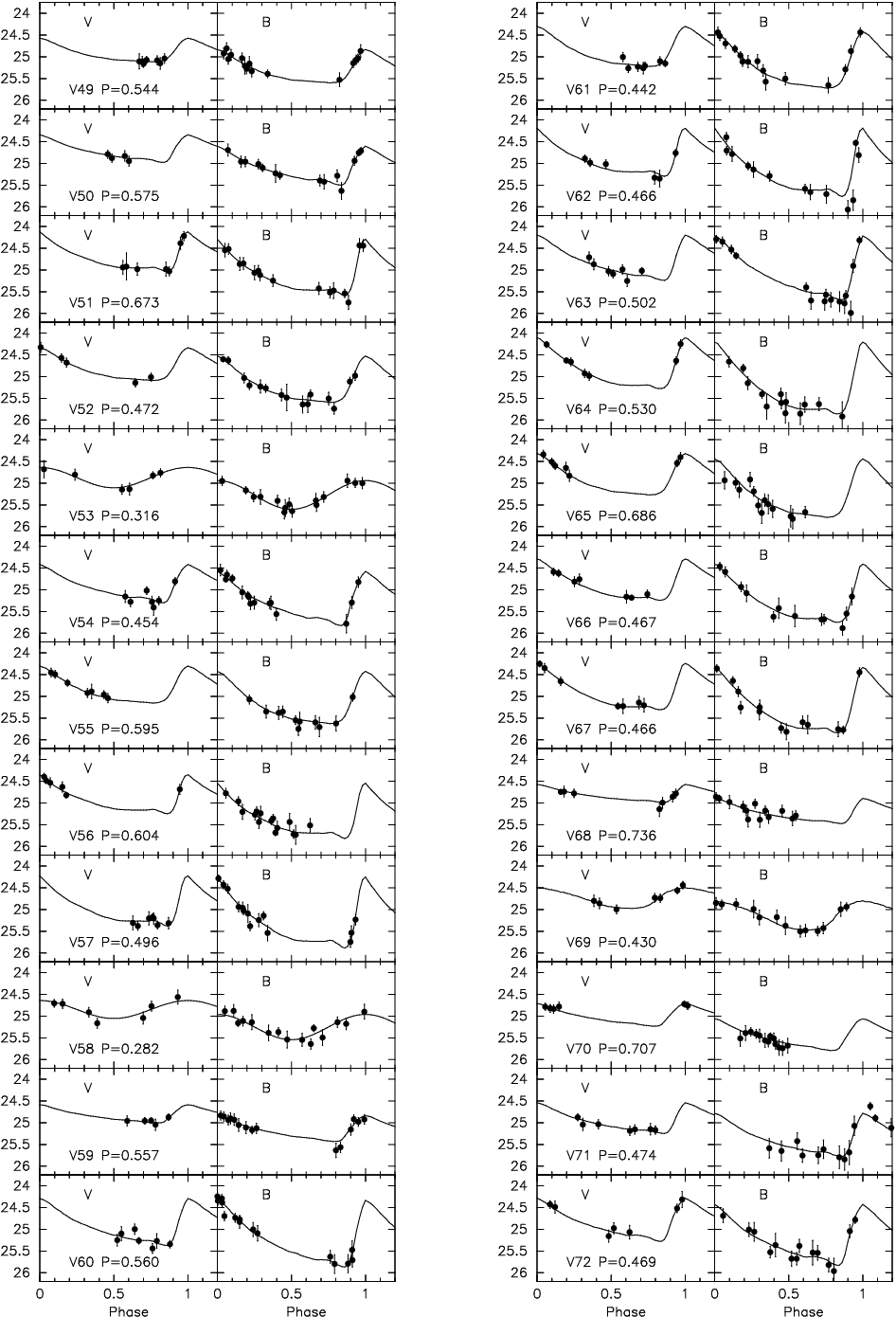,height=7.00in,width=5.00in}}
  \caption{And~II variable star light curves.  The observations are shown
   as filled circles and the fitted templates are displayed as curves.}
  \label{Fig03}
\end{figure*}

\begin{figure*}[t]
  \figurenum{3 cont}
  \centerline{\psfig{figure=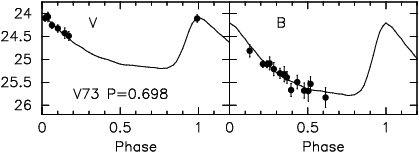,height=1.25in,width=3.00in}}
  \caption{And~II variable star light curves.  The observations are shown
   as filled circles and the fitted templates are displayed as curves.}
  \label{Fig03}
\end{figure*}

\section{And~II Anomalous Cepheid} 

We were able to find one AC, V14, in the field-of-view of the WFPC2.  This 
follows the trend of at least one AC being found in every dSph galaxy surveyed 
for variable stars.  ACs are believed to be either stars that have increased 
mass due to mass transfer in a binary system (Renzini, Mengel, \& Sweigart 
1977) or stars from an intermediate age population (Demarque \& Hirshfeld 
1975; Norris \& Zinn 1975) although metal-poor abundances are required in 
both cases.  The main way to recognize ACs is by their higher 
luminosity when compared to the RRLs in a system, typically 0.5-2.0~mag 
brighter than the HB.\@  

The ACs follow a period-luminosity relation different than that for classical 
Cepheids or Population~II Cepheids, hence the name ``anomalous."  In Paper~I 
we revised the period-luminosity relations for the ACs (see \S4.1 in 
Paper~I).\@  Using the data from Table~2 and given $(m-M)_0 = 24.17\pm0.06$ and 
$E(\bv)=0.06\pm0.01$ for And~II from DACS00, we find $A_V = 3.1 E(\bv) = 
0.19\pm0.03$ resulting in $M_V = -0.75$~mag and $M_B = -0.43$~mag for V14.  
This places the AC in And~II with the other known ACs (see Table~4 of Paper~I) 
in the period-luminosity diagram in Figure~4.  V14 falls nicely along the 
first-overtone AC line in both $M_B$ and $M_V$.\@  

We note that the light curve shape of V14 is quite asymmetric which would 
lead one to think it is pulsating in the fundamental mode.  Yet, it is 
clearly a first-overtone mode pulsator as seen in Figure~4.  This reinforces 
the view we expressed in Paper~I that the shape of the light curve isn't as 
clear an indicator of pulsation mode for an AC as is its location in a plot of 
the absolute magnitude versus the logarithm of the period.  Given the 
$B$ and $V$ amplitudes for V14 in Table~2, this star also falls nicely among 
the other first-overtone mode AC pulsators in the period-amplitude diagram 
shown as Fig.~7 of Paper~I.\@

\begin{figure*}[t]
  \centerline{\psfig{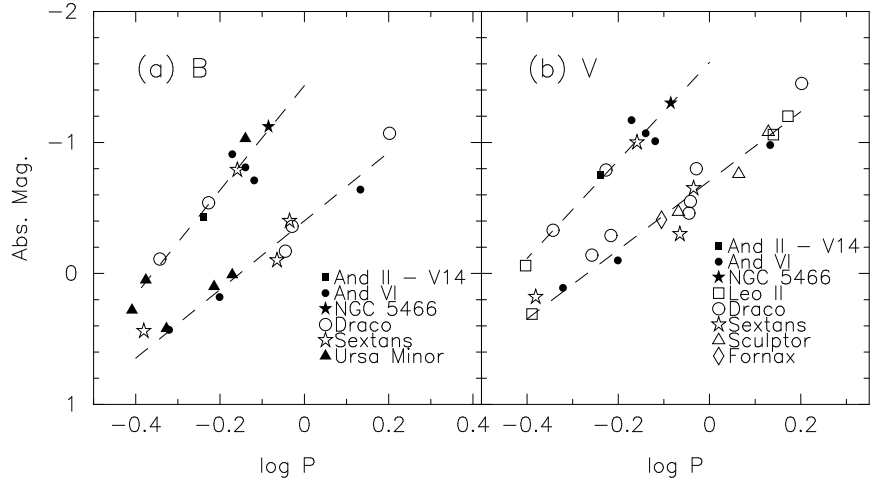}}
  \caption{Absolute magnitude versus period diagrams for anomalous 
   Cepheids.  Only those anomalous Cepheids with quality light curves 
   are shown for the (a) $B$ and (b) $V$ filters.  The dashed lines 
   represent the fits to the fundamental and first-overtone anomalous 
   Cepheids taken from Paper~I.\@  The data for the dwarf spheroidal 
   galaxies besides And~II were taken from Paper~I.\@  The And~II 
   AC V14 is shown as the filled square.}
  \label{Fig04}
\end{figure*}

\section{And~II RR Lyrae} 

RRLs are useful when examining the stellar populations of a system.  
Their periods, amplitudes, and magnitudes are probes of the 
distance, metallicity, and age of the system in which they belong.  We found 
72 RRLs in our field-of-view of And~II, with 64 pulsating in the fundamental 
mode (RRab) and eight pulsating in the first-overtone mode (RRc).  Their mere 
presence indicates that there are stellar populations of age $> 10$~Gyr in 
And~II.\@  Due to the low amplitudes of the RRc stars and the uncertainties in 
our photometry, it is likely that we did not detect all of this type of 
RRLs.  We compare the period distribution of the RRL population of And~II to 
those of other dSph galaxies in Figure~5 using the sources listed in 
Table~6 of Paper~I and the values derived above for And~II.  The mean 
metallicity increases from 
the top, Ursa Minor, to the bottom, Fornax.  It should be noted that although 
the mean metallicity of Fornax is $\langle {\rm [Fe/H]} \rangle = -1.3$~dex 
(Mateo 1998), the RRLs may originate from the metal-poor end of Fornax's 
metallicity distribution (Bersier \& Wood 2002).  Looking at the RRab stars, 
there appears to be an overall trend of these stars shifting toward shorter 
periods as the metallicity increases.  And~II shows more of a spread in period 
among its RRab stars as compared to the other dSph galaxies.  It is uncertain 
how much of this distribution may be due to the spread in metallicity that is 
known to exist in this dSph galaxy ($\sigma_{\rm int, [Fe/H]} \approx 
0.36$~dex, DACS00). 

The mean periods for the RRLs are $\langle P_{ab} \rangle = 0.571$~day 
and $\langle P_c \rangle = 0.363$~day.  In comparison to the period-metallicity 
relationship defined by the Galactic globular clusters, the mean period for 
the RRab stars is consistent with the mean metallicity in And~II as determined 
from the red giant branch.  For the RRc stars, however, the mean period 
is slightly longer than that expected if these stars have the same mean 
metallicity as the red giants.  As discussed above, the RRc mean period 
may be unreliable as it is possible we are missing many of these stars.  
Nevertheless, the ratio of RRc stars to the total number of RRLs, 
$N_c/N_{RR}$, is 0.11, consistent with the relatively high mean 
metallicity of And~II.\@  If we place And~II in Table~6 of Paper~I using 
the mean metallicity from the red giant branch, it follows the trend in 
which the mean period for the RRab stars decreases as the metallicity 
increases.  To illustrate this, we reproduce the metallicity versus mean 
RRab period diagram of Paper~I (see Fig.~10 of that paper) in Figure~6, but now 
include And~II.\@  With the exception of $\omega$~Centauri and M2, there is 
a clear trend of increasing mean period with decreasing ${\rm [Fe/H]}$ for 
the Galactic globular clusters.  As noted in Paper~I, the mean periods of the 
RRab stars in the dSph galaxies fill in the gap between the metal-rich and 
metal-poor Galactic globular clusters.  And~II clearly also follows this 
trend -- it lies in Figure~6 among the Galactic globular clusters which have 
similar metallicities to the And~II mean abundance.

\begin{figure*}[t]
 \centerline{\psfig{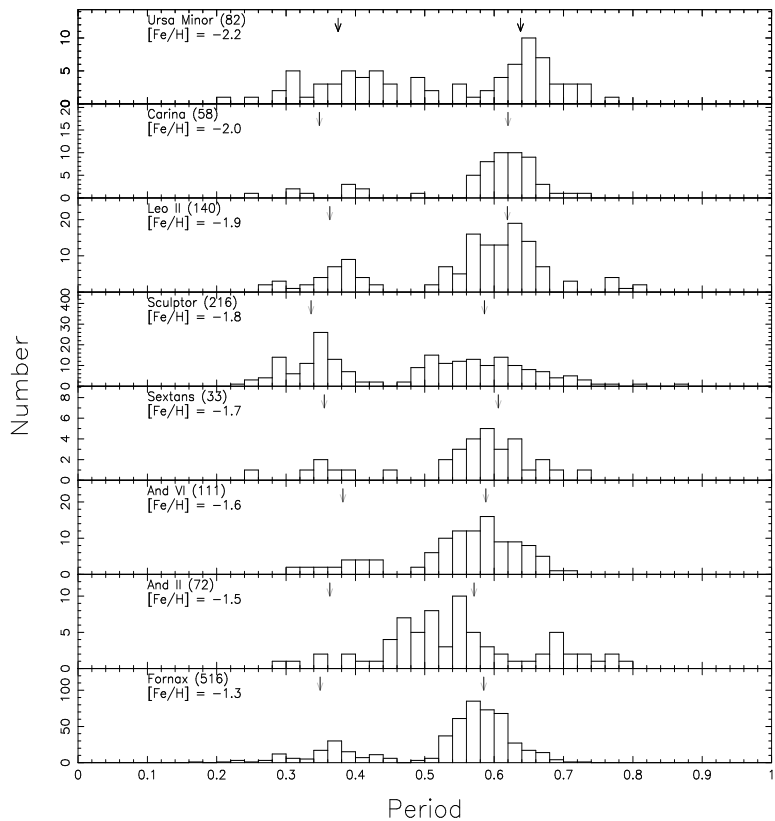}}
 \caption{Period distribution plots for the RR~Lyrae stars in dwarf 
   spheroidal galaxies.  The plots are arranged so that the mean 
   metallicity of the dwarf spheroidal galaxy increases from the top down.  
   For each dwarf spheroidal galaxy, the vertical arrows indicate the mean 
   periods for the RRc (left) and RRab (right) stars.}
 \label{Fig05}
\end{figure*}

\begin{figure*}[t]
 \centerline{\psfig{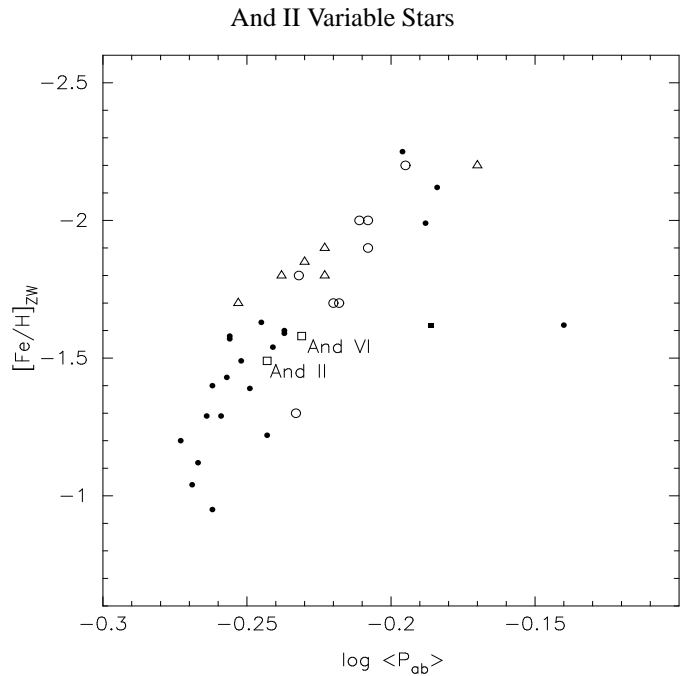}}
 \caption{Mean period for the RRab stars versus the metallicity of the 
   parent system.  The Andromeda dwarf spheroidal galaxies are shown as 
   open squares.  The Galactic dwarf spheroidal galaxies are indicated 
   by open circles.  Galactic globular clusters with at least 15 RRab 
   stars are shown as filled circles, along with $\omega$~Centauri 
   (filled square).  Large Magellanic Cloud globular clusters with a 
   minimum of 15 RRab stars are shown as open triangles.}
 \label{Fig06}
\end{figure*}

We find the mean magnitude of the And~II RRLs to be 
$\langle V \rangle = 24.87\pm0.03$~mag, where the uncertainty is the aperture 
correction uncertainty, the photometry zeropoint uncertainty, and the 
spline-fitting uncertainty added in quadrature to the standard error of the 
mean.  This matches the estimate of the level of the HB by DACS00, 
$V_{\rm HB} = 24.93\pm0.03$~mag based on the mean magnitude of the HB stars.  
We used the same equation as DACS00 to estimate the absolute magnitude of the 
RRLs, 

\begin{equation} 
M_{V,{\rm RR}} = 0.17{\rm [Fe/H]} + 0.82
\end{equation} 

\noindent
from Lee, Demarque, \& Zinn (1990).  We find $M_{V,{\rm RR}}=+0.57$ and 
$A_V=3.1 E(\bv)=0.19\pm0.03$ given $\langle {\rm [Fe/H]} \rangle = 
-1.49\pm0.11$ and $E(\bv)=0.06\pm0.01$ from DACS00.  As a result, we estimate 
the distance to And~II to be $665\pm20$~kpc, which matches the 
estimate made by DACS00 of $680\pm20$~kpc.

\subsection{RR~Lyrae Period-Amplitude Diagram}

The period-amplitude diagram provides an excellent diagnostic of the properties 
of the RRLs since neither of these quantities is dependent on distance or 
reddening.  It is generally true that the position of RRLs in a 
period-amplitude diagram is dependent on its metallicity (Sandage 1993b).  
Still, there are other factors besides metallicity, such as evolution and 
age, that may shift a RRL's position in the diagram.  There is a clear 
division between the shorter period, smaller amplitude RRc stars and the 
RRab stars in the And~II period-amplitude diagram (Figure~7).  
A wide spread in $\log\,P$ is seen among the RRab stars for a given $A_B$, 
that may or may not be due to the spread in ${\rm [Fe/H]}$ for And~II (DACS00).  

There may also be some aliasing in the periods.  As with And~VI (Paper~I), 
we used the period-finding routines created by Dr.\ Andrew Layden 
(Layden \& Sarajedini 2000 and references therein) to 
find the best period for each variable.  We also asked Dr.\ Gisella Clementini 
to determine what periods she would find for a small number of our variable 
stars using her period-finding program GRATIS (GRaphical Analyzer of TIme 
Series, see Clementini et al.\ 2000, 2003).  The periods she found matched 
those determined from Andrew Layden's program to $\pm 0.001$~day.  For And~VI 
we were able to make use of the period-amplitude diagram to see if the 
period adopted for each star was reasonable.  However, the red giant branch 
population of And~II shows a large abundance dispersion which may well 
also occur among the RRLs.  This limits the usefulness of the period-amplitude 
diagram in determining the most appropriate period.  Therefore, we have adopted 
for each star the best-fit period that was returned from Layden's 
period-fitting program with no further screening.  

\begin{figure*}[t]
 \centerline{\psfig{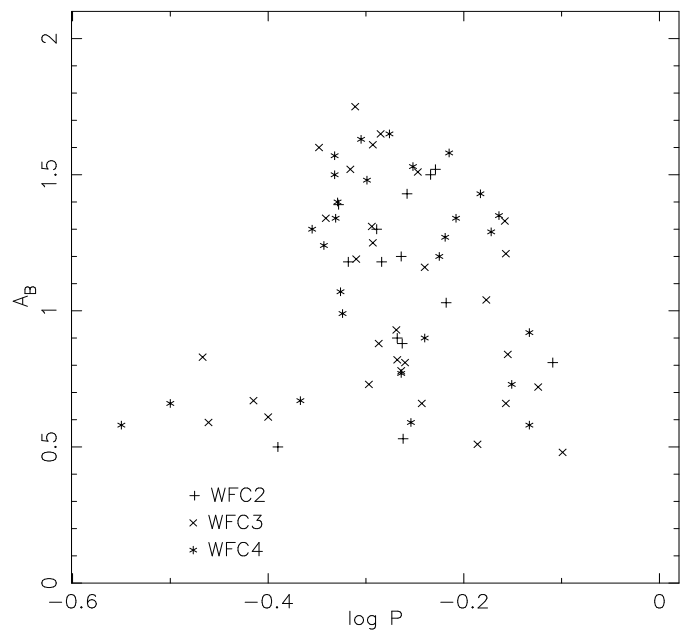}}
 \caption{Period-amplitude diagram for the RR~Lyrae stars in And~II.\@  The 
   amplitudes shown are for the $B$~filter.}
 \label{Fig07}
\end{figure*}

Sandage, Katem, \& Sandage (1981) detected a shift in period between the 
RRLs in M3 and M15 in the period-amplitude diagram.  Sandage (1981; 1982a,b) 
later found that this period shift was dependent on the metallicity of 
a cluster where metal-poor clusters tend to have longer periods for a 
given amplitude when compared to metal-rich clusters.  Using M3 as the 
fiducial cluster with zero period shift, Sandage (1982a,b) found 

\begin{equation} 
\Delta\log\,P = -(0.129A_B + 0.088 + \log\,P).
\end{equation}

\noindent 
Sandage found that more metal-rich clusters have 
$\langle\Delta\log\,P\rangle \ge -0.01$, while the more metal-poor clusters 
have $\langle\Delta\log\,P\rangle \le -0.05$.  Between these two values, a 
gap exists where few RRLs are found.  On the other hand, as discussed in 
Paper~I (see \S5.2), dSph galaxies have been shown to contain RRLs within the 
$-0.05 \le \Delta\log\,P \le -0.01$ range (Sextans:  Mateo et al.\ 1995; 
Leo~II:  Siegel \& Majewski 2000; Andromeda~VI:  Paper~I).\@  We list in 
column~10 of Table~2 the $\Delta\log\,P$ values for the RRab stars of 
And~II.\@  In Figure~8 we plot these values against $\log\,P$ for each RRab 
star.  It is clear in the figure that a number of And~II RRab stars are found 
in-between $-0.05 \le \Delta\log\,P \le -0.01$.  We find that a majority of 
the RRab stars are found in the metal-rich region of the diagram 
($\Delta(\log\,P) > -0.01$).  This is 
consistent with the mean metallicity of And~II.\@  When comparing And~II to 
And~VI (see Fig.~12 of Paper~I), we see that there is a larger range in 
$\Delta\log\,P$ values in And~II than in And~VI.\@  This suggests that 
there is a sizable abundance spread among the And~II RRLs, a result that is 
perhaps not unexpected given the large abundance dispersion among the red 
giants in this dSph galaxy.  The mean period shift for And~II is 
0.01, which means that the RRab in And~II are slightly more metal-rich than 
M3.  Using 

\begin{equation} 
\Delta\log\,P = 0.116 {\rm [Fe/H]} + 0.173 
\end{equation} 

\noindent 
from Sandage (1982a), we find ${\rm [Fe/H]} = -1.41$ for And~II.\@  This 
estimate of the mean metal abundance of the And~II RRLs is consistent 
with the mean abundance for And~II red giants found by DACS00, $\langle 
{\rm [Fe/H]} \rangle = -1.49$.

\begin{figure*}[t]
 \centerline{\psfig{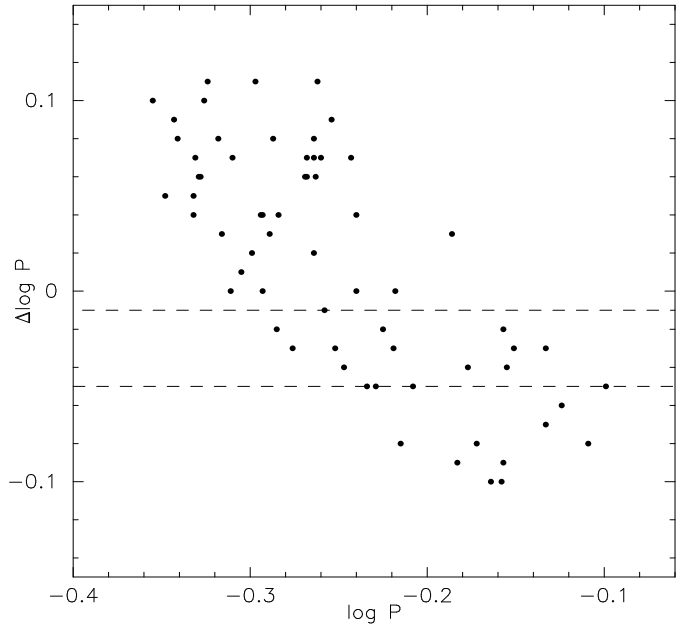}}
 \caption{Period shift versus period for the RRab in And~II.\@  The 
  dashed lines represent the zone in which few Galactic globular cluster 
  RRab stars are found.}
 \label{Fig08}
\end{figure*}

\subsection{Metallicity Estimates from RR~Lyrae}

Alcock et al.\ (2000) developed a relation between the metallicity of a RRab 
star and its period and $V$-band amplitude.  This was done in order to 
estimate the metallicity of the RRab stars in the Large Magellanic Cloud.  
Using the RRab stars in M3, M5, and M15, Alcock et al.\ found, 

\begin{equation} 
{\rm [Fe/H]}_{\rm ZW} = -8.85(\log\,P_{ab} + 0.15A_V) - 2.60{\rm ,} 
\end{equation}

\noindent
where ZW refers to the Zinn \& West (1984) scale.  The metallicity of the 
RRab stars in those systems was predicted with an accuracy of 
$\sigma_{\rm [Fe/H]} = 0.31$ per star.  For the LMC, Alcock et al.\ found 
good agreement between the median metallicity from their relation and those 
from previous estimates.  We applied this relation to our And~VI RRLs 
(see \S5.3 of Paper~I) and found the mean of the RRL metallicity distribution 
generated from this relation to agree well with the mean metallicity for the 
red giants determined by Armandroff, Jacoby, \& Davies (1999) using the red 
giant branch mean $V-I$ color.  The width of the metallicity distribution, 
however, was essentially identical to that expected from the uncertainty 
($\sim 0.3$~dex) in the individual abundance determinations using this 
relation.  Consequently, from these data, there is no evidence to support 
(or rule out) an abundance spread among the And~VI RRLs.

\begin{figure*}[t]
 \centerline{\psfig{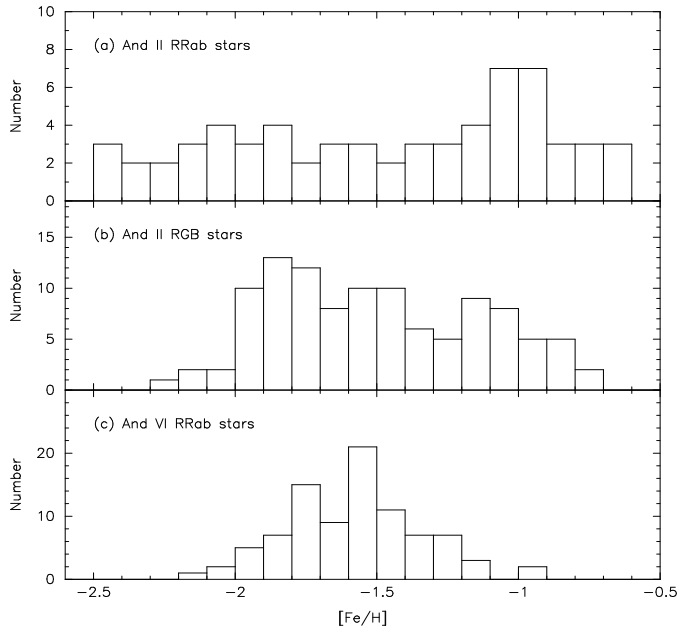}}
 \caption{[Fe/H] distribution plots for (a) the RRab stars in And~II, 
  (b) the RGB stars in And~II from DACS00, and (c) the RRab stars in 
  And~VI from Pritzl et al.\ (2002).  The individual [Fe/H] values for 
  the RRab stars were calculated using Eq.\ 2 in Alcock et al.\ (2000) 
  and have an individual uncertainty of $\sim0.3$~dex.  The abundances 
  for the RGB stars have errors of $\sim0.08$~dex (DACS00).  Although 
  there is no reason to expect it to be true, the distributions for the 
  And~II RRab stars and RGB stars are comparable.  Their widths are much 
  larger than that found for the And~VI RRab stars ($\sigma\sim0.3$).}
 \label{Fig09}
\end{figure*}

We applied this equation to the RRab in And~II and list the individual 
metallicities in column~11 in Table~2.  Figure~9a shows the distribution 
of the individual ${\rm [Fe/H]}$ from the RRab in And~II.\@  There is a wide 
range among the metallicity values similar to the histogram of metallicity 
values derived from red giant colors for And~II as found by DACS00 and shown 
in Figure~9b.  We also plot in Figure~9c the [Fe/H] distribution from the
RRab stars in And~VI (Paper~1).  This figure emphasizes 
the differences between the metallicity distributions.  Where the width of 
the [Fe/H] distribution for the And~VI RRab stars was on the order of the 
uncertainty of the Alcock et al.\ equation ($\sigma\sim0.3$), the width for 
the And~II stars is clearly larger than this.

Due to the wide spread in the metallicities, we did not attempt to fit a 
gaussian to the distribution as was done for And~VI in Paper~I.\@  However, 
we find the mean to be ${\rm [Fe/H]}=-1.46$ and the median to be 
${\rm [Fe/H]}=-1.36$.  These both agree well with the mean metallicity found 
by DACS00 using the colors of the red giant branch stars, $\langle 
{\rm [Fe/H]} \rangle =-1.49\pm0.11$.  It appears that the mean metallicity 
values resulting from RRLs and red giants in And~II are consistent.  
Even though the metallicity distributions for the RGB and RRab stars are 
comparable, there is no reason why one should expect this.  The mapping from 
the RGB to the instability strip is a complex function of (at least) 
metallicity, age distributions, and evolution within the instability strip 
 as well as whatever parameters control 
mass-loss.  Most And~II RGB stars do not become a RRL, which is clear 
from looking at the stellar distribution along the HB.  Nevertheless, it is 
intriguing that the range of abundances from the RGB and RRab stars is 
similar and they both show a peak in the distribution near the metal-rich end.

Studying cluster and field RRLs with a wide range of metallicities, Sandage 
(1993a) related the metallicities to the average periods of RRLs through 
the relations 

\begin{equation}
{\rm [Fe/H]}_{\rm ZW} = (-\log\,\langle\,P_{ab}\rangle - 0.389)/0.092 
\end{equation} 
\begin{equation} 
{\rm [Fe/H]}_{\rm ZW} = (-\log\,\langle\,P_{c}\rangle - 0.670)/0.119 {\rm .}
\end{equation}

\noindent
Siegel \& Majewski (2000) and Cseresnjes (2001) found that these relations 
applied well to the RRLs in dSph galaxies.  We found in Paper~I that the 
RRab estimate for And~VI was within the combined errors of the estimate 
and that of Armandroff, Jacoby, \& Davies (1999).  The RRc abundance estimate, 
however, was lower than expected from the RRab and red giant branch 
stars, perhaps as a consequence of incompleteness in the sample of RRc stars, 
especially at shorter periods.  The mean periods for the RRLs in And~II 
are $\langle P_{ab} \rangle = 0.571\pm0.005$~day and $\langle P_c \rangle = 
0.363\pm0.005$~day giving ${\rm [Fe/H]}_{\rm ZW} = -1.58\pm0.04$ (internal 
error) for the RRab and ${\rm [Fe/H]}_{\rm ZW} = -1.93\pm0.05$ (internal 
error) for the RRc.  The metallicity estimate from the RRab stars, although 
slightly lower than the estimate made by DACS00, is within the combined 
errors.  Once again, however, the estimate from the RRc stars is lower than 
that for the RRab and the red giant branch stars.  It is possible that we 
have not detected all the RRc stars, especially at shorter periods, 
resulting in an inappropriate mean period (see \S5).

\section{The Frequency of Anomalous Cepheids in dSph Galaxies} 

Mateo, Fischer, \& Krzeminski (1995) examined the specific frequency
of ACs (the number of ACs per $10^5 L_{V,\odot}$) in the Galactic dSph
galaxies and found a strong correlation between the specific frequency of
ACs and the total luminosities of the parent galaxies: 
low luminosity dSph galaxies have higher specific frequencies of ACs.
Interpretation of this correlation, however, is complicated. For example, 
Draco, Ursa~Minor, and Carina all have
similar luminosities and similar AC specific frequencies, but they have
very different stellar populations. Draco and Ursa Minor are dominated 
by old stars and thus, in these particular dSphs, the ACs must have their
origin in the binary mass transfer mechanism. 
On the other hand, Carina has a strong intermediate-age population and
consequently, at least some of the ACs in this dSph may originate as
single younger stars, rather than from (old) mass transfer binaries. 
Given these differences, there is no obvious reason why Carina should 
end up with a similar specific frequency to Draco and Ursa Minor. Similarly,
Leo~I has a dominant and relatively young intermediate-age metal-poor
population (Gallart et al.\ 1999), yet its specific frequency of ACs 
is notably lower than that of Draco, Ursa Minor and Carina. Mateo et al.\ 
(1995) noted that dSph luminosities are strongly correlated with mean
abundance (lower abundances for lower luminosities) and that this might
be the underlying cause of the AC specific frequency - luminosity 
correlation: regardless of the origin of the ACs, they occur more
frequently in metal-poorer environments.

\begin{figure*}[t]
 \centerline{\psfig{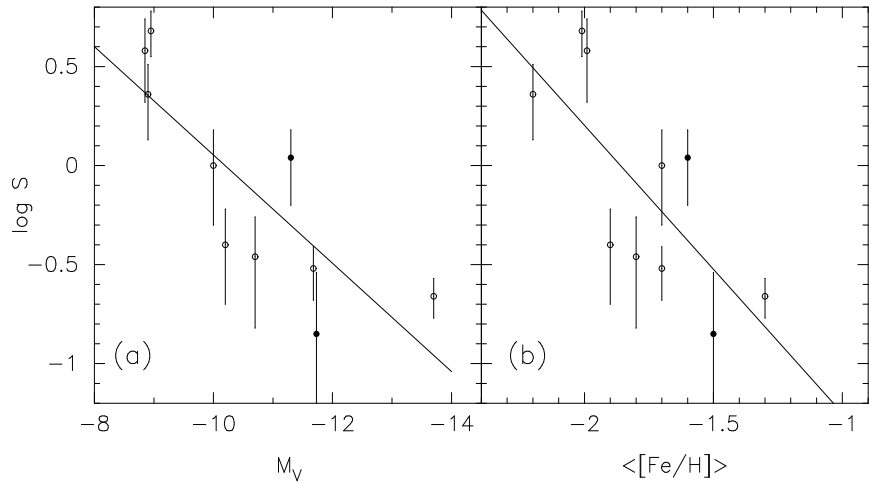}}
 \caption{Specific frequency of anomalous Cepheids in dwarf spheroidal 
  galaxies as a function of (a) the $M_V$ and (b) the mean [Fe/H] of the 
  galaxy.  The Galactic dwarf spheroidal galaxies are shown as open circles 
  and the M31 ones are shown as filled circles.  In both plots, there are 
  clear trends that the Galactic dwarf spheroidal galaxies follow as 
  shown by the least squares fit lines fit to their data.  Both And~II and 
  And~VI follow these trends.}
 \label{Fig10}
\end{figure*}

There are now several more surveys for variable stars
in dSph galaxies, including our surveys of the M31 dSph galaxies,
than were available to Mateo et al.\ (1995). This may help disentangle the
independent parameter(s) that drive the AC frequencies. In Table~5, we
present revised values for Mateo et al.\ Table~7; we include the results
from the latest surveys. For example, Dall'Ora et al.\ (2003) have
identified 15 ACs in Carina, almost twice as many as previously known.
In Table~5, Column~1 lists the dSph galaxy and column~2 is
the mean [Fe/H]. The absolute magnitude of the galaxy is in column~3,
and the structural parameters for each galaxy are given in columns~4--6
as the position angle measured from the North to the East, the ratio of
the semi-minor axis to the semi-major axis, and the exponential scale length.
The latter parameters are taken from the references given in the notes
to the table.
The percentage of the dSph luminosity surveyed and the number of ACs found
are listed in columns~7 and 8, respectively. The specific frequencies of
the ACs are then given in column~9, along with the logarithms of the specific
frequencies and their uncertainties in column~10. $S$ is defined as
the number of ACs per $10^5 L_{V,\odot}$ and was calculated for only the
fraction of the dSph's luminosity that was surveyed for variable stars.
The uncertainties in the $\log\,S$ values were determined by assuming
that the number of observed ACs is subject to Poisson statistics.

In order to calculate the luminosity surveyed in And~VI and And~II, we
first determined the orientation and location of the region covered by 
the WFPC2 images relative to the galaxy. We then calculated the intensity of
the WFPC2 overlap region relative to the entire galaxy by approximating
the surface brightness distribution of the galaxy by the relation 
$I=I_0 e^{-(r/r_0)^n}$ (Sersic 1968), and making the appropriate allowance
for the ellipticities. 
The $V$-band surface brightness profile parameters were taken from Caldwell 
et al.\ (1992) for And~II and from Caldwell (1999) for And~VI\@. From the 
luminosity fractions just calculated and the total visual magnitudes of the 
galaxies (column~3 of Table 5) we then derive the ($V$-band) luminosity 
encompassed by our images of each galaxy to which the AC surveys apply.  
The same process was applied for the Fornax dSph galaxy according to the 
survey area of the Bersier \& Wood (2002) survey. 

Using the data from Table 5,
the logarithm of the specific frequency of ACs is plotted against $M_{V}$
for each galaxy in Fig.\ 10a, while in Fig.\ 10b, the logarithm 
of the specific frequency is plotted against the mean [Fe/H] for each dwarf. 
The Galactic dSph galaxies are shown as open circles and the M31 dSph 
galaxies as filled circles. In each plot the solid line is an unweighted
least squares fit to the Galactic dSph data only. 
We confirm in Fig.\ 10a that a $\log\,S$ -- (visual) luminosity correlation 
exists (correlation coefficient is 0.86) for the Galactic dSph galaxies.  
The correlation between $\log\,S$
and mean abundance shown in Fig.\ 10b is not as strong (correlation 
coefficient 0.74) but it is nevertheless significant -- a correlation test
shows that there is only a 4\% probability (2.3-sigma) that the distribution 
of $\log\,S$ vs [Fe/H] arises by chance. This supports the contention
of Mateo et al.\ (1995) that the mean abundance of dSph galaxies may be 
the underlying factor that governs the frequency of occurrence of AC 
variables, at least for the Galactic dSphs.

The addition of data for the M31 dSphs then allows us to investigate,
at least in a preliminary way, whether these correlations apply also
to the AC populations in dSph galaxies other than those of the Galaxy.
The points for And~II and And~VI in Figs.\ 10a and 10b, indicate that these
galaxies do appear to follow the relations as defined by the Galactic dSphs,
but results for additional M31 dSphs are required to strengthen this
conclusion. We will return to this question in our third paper in this
series, where we will explore the variable star population, including
ACs, of the M31 companions And~I and And~III.

\section{Summary \& Conclusions} 

Although it is one of the more metal-rich dSph galaxies, the properties of the 
variable stars in And~II agree well with the trends defined by other dSph 
galaxies.  We have found 73 variable stars in 
And~II using HST/WFPC2 observations, one of which is an AC.\@  The AC is found 
to have a period, absolute magnitude, and amplitude consistent with other ACs.  
The 72 RRLs were made up of 
64 RRab stars and 8 RRc stars with mean periods of $\langle P_{ab} \rangle 
= 0.571\pm0.005$~day and $\langle P_c \rangle = 0.382\pm0.005$~day.  The 
mean $V$ magnitude of the RRLs was found to be $24.87\pm0.03$~mag resulting 
in a distance of $665\pm20$~kpc for And~II on the Lee, Demarque, \& Zinn 
(1990) distance scale.  

Using relations from Sandage (1982a, 1993a) and Alcock et al.\ (2000), we find 
that the properties of the RRab stars in And~II yield estimates 
of the mean abundance of these stars that are in good agreement with the 
mean metallicity determined from both the colors (DACS00) and the line 
strengths (C\^{o}t\'{e} et al.\ 1999) of And~II's red giants.  With these 
results And~II then follows the general trend seen in the Galactic 
globular clusters and Galactic dSphs in which the mean RRab period decreases 
with increasing metallicity.  In addition, the And~II RRLs show 
considerable spread in the period-amplitude diagram.  We take this as an 
indication that there is a sizable abundance spread among these stars and 
have attempted to quantify this spread through use of the Alcock et al.\ 
(2000) relation.  This yields an abundance distribution for the RRL which 
is considerably broader than that expected from the errors in the individual 
abundance determinations and contrasts with the case for And~VI (Paper~I) 
where there was no evidence for any intrinsic abundance spread among the RRL 
stars.  The And~II RRL abundance distribution is approximately uniform and 
has some features in common with the large metallicity spread derived by 
C\^{o}t\'{e} et al.\ (1999) and DACS00 for the red giant branch stars 
(see Fig.~9).

We have investigated the existence of a relationship between the specific 
frequency of ACs and mean metallicities for dSph galaxies.  As originally 
suggested by Mateo et al.\ (1995), we find for the Galactic dSphs there is 
a clear trend for higher specific frequencies at lower abundances.  The M31 
dSph galaxies And~II and And~VI also appear to follow this trend.  However, 
more information on the frequency of occurrence of ACs in additional M31 dSph 
galaxies is required before we can fully compare the two sets of dSphs in 
this parameter plane.  

\acknowledgements

This research was supported in part by NASA through grant number GO-08272 
from the Space Telescope Science Institute, which is operated by AURA, Inc., 
under NASA contract NAS 5-26555.

We would like to thank Peter Stetson for graciously sharing his PSFs for 
the WFPC2 and for the use of his data reduction programs.  Thanks to Andrew 
Layden for the use of his light curve analysis programs.  We would also like 
to thank Gisella Clementini for testing our period determinations.  Thank you 
to Mario Mateo for his insightful referee comments and suggestions.

\begin{deluxetable}{ccccc} 
\tablewidth{0pc} 
\tablecaption{Photometric Differences \label{tbl-1}} 
\tablehead{
\colhead{Chip} & \colhead{$\Delta V$} & \colhead{\# $V$ Stars} & 
\colhead{$\Delta B$} & \colhead{\# $B$ Stars} \\ 
 & \colhead{$21.5<V<25.5$} & & \colhead{$23.0<B<26.0$} & \\
          } 
\startdata 
WFC2 & --0.050 $\pm$ 0.002 & 302 & --0.100 $\pm$ 0.003 & 261 \\
WFC3 & --0.076 $\pm$ 0.002 & 487 & --0.094 $\pm$ 0.003 & 435 \\ 
WFC4 & --0.070 $\pm$ 0.002 & 405 & --0.062 $\pm$ 0.003 & 360 \\ 
\enddata
\tablecomments{difference = magnitude in present study -- Da Costa et al.\ 
(2000) magnitude} 
\end{deluxetable}

\begin{deluxetable}{cccccccccccc} 
\tablewidth{0pc} 
\tabletypesize{\scriptsize} 
\tablecaption{Light Curve Properties \label{tbl-2}} 
\tablehead{
\colhead{ID} & \colhead{RA (2000)} & \colhead{Dec (2000)} & \colhead{Period} & 
\colhead{$\langle V \rangle$} & 
\colhead{$\langle B \rangle$} & \colhead{$(B-V)_{\rm mag}$}  & 
\colhead{$A_V$} & \colhead{$A_B$} & \colhead{$\Delta\log\,P$} & 
\colhead{[Fe/H]} & \colhead{Classification} \\ 
\colhead{(1)} & \colhead{(2)} & \colhead{(3)} & \colhead{(4)} & 
\colhead{(5)} & \colhead{(6)} & \colhead{(7)} & \colhead{(8)} & 
\colhead{(9)} & \colhead{(10)} & \colhead{(11)} & \colhead{(12)}  
          } 
\startdata 
V01  & 1:16:20.49 & 33:26:11.8 & 0.407 & 24.808 & 25.211 & 0.410 & 0.35 & 0.50 & \nodata & \nodata & c \\
V02  & 1:16:20.25 & 33:26:13.4 & 0.546 & 24.963 & 25.343 & 0.396 & 0.62 & 0.88 &  0.06 & -1.10 & ab \\ 
V03  & 1:16:21.73 & 33:26:13.9 & 0.520 & 24.800 & 25.178 & 0.412 & 0.83 & 1.18 &  0.04 & -1.19 & ab \\ 
V04  & 1:16:19.88 & 33:26:35.6 & 0.540 & 24.812 & 25.144 & 0.350 & 0.64 & 0.90 &  0.06 & -1.08 & ab \\ 
V05  & 1:16:20.52 & 33:26:37.0 & 0.583 & 24.926 & 25.274 & 0.403 & 1.06 & 1.50 & -0.05 & -1.93 & ab \\ 
V06  & 1:16:21.61 & 33:26:34.0 & 0.590 & 24.870 & 25.267 & 0.447 & 1.08 & 1.52 & -0.05 & -2.01 & ab \\ 
V07  & 1:16:21.48 & 33:26:36.1 & 0.481 & 24.781 & 25.154 & 0.408 & 0.83 & 1.17 &  0.08 & -0.89 & ab \\ 
V08  & 1:16:22.27 & 33:26:34.6 & 0.552 & 24.812 & 25.246 & 0.477 & 1.02 & 1.43 & -0.01 & -1.67 & ab \\ 
V09  & 1:16:19.54 & 33:26:54.4 & 0.547 & 24.939 & 25.401 & 0.468 & 0.37 & 0.53 &  0.11 & -0.77 & ab \\ 
V10  & 1:16:23.03 & 33:26:40.6 & 0.605 & 24.870 & 25.288 & 0.444 & 0.73 & 1.03 &  0.00 & -1.64 & ab \\ 
V11  & 1:16:19.30 & 33:26:57.1 & 0.778 & 24.929 & 25.412 & 0.496 & 0.58 & 0.81 & -0.08 & -2.41 & ab \\ 
V12  & 1:16:20.68 & 33:26:54.4 & 0.514 & 24.901 & 25.370 & 0.504 & 0.92 & 1.30 &  0.03 & -1.26 & ab \\ 
V13  & 1:16:20.11 & 33:27:01.4 & 0.544 & 24.855 & 25.271 & 0.453 & 0.85 & 1.20 &  0.02 & -1.39 & ab \\ 
V14  & 1:16:20.57 & 33:27:01.5 & 0.578 & 23.608 & 23.990 & 0.419 & 0.91 & 1.29 & \nodata & \nodata & AC \\ 
V15  & 1:16:20.21 & 33:27:21.7 & 0.470 & 24.889 & 25.311 & 0.463 & 0.99 & 1.39 &  0.06 & -1.01 & ab \\ 
V16  & 1:16:23.80 & 33:26:38.1 & 0.346 & 24.848 & 25.086 & 0.248 & 0.42 & 0.59 & \nodata & \nodata & c \\ 
V17  & 1:16:23.55 & 33:26:14.9 & 0.575 & 24.944 & 25.371 & 0.456 & 0.82 & 1.16 &  0.00 & -1.57 & ab \\ 
V18  & 1:16:24.20 & 33:26:35.1 & 0.700 & 24.933 & 25.201 & 0.285 & 0.59 & 0.84 & -0.04 & -2.01 & ab \\ 
V19  & 1:16:23.55 & 33:26:10.0 & 0.489 & 24.800 & 25.096 & 0.372 & 1.24 & 1.75 &  0.00 & -1.50 & ab \\ 
V20  & 1:16:23.88 & 33:25:54.6 & 0.508 & 24.941 & 25.276 & 0.371 & 0.92 & 1.31 &  0.04 & -1.22 & ab \\ 
V21  & 1:16:24.16 & 33:26:02.1 & 0.456 & 24.821 & 25.156 & 0.373 & 0.95 & 1.34 &  0.08 & -0.84 & ab \\ 
V22  & 1:16:25.81 & 33:26:56.4 & 0.398 & 24.850 & 25.133 & 0.293 & 0.43 & 0.61 & \nodata & \nodata & c \\ 
V23  & 1:16:25.12 & 33:26:29.2 & 0.540 & 24.845 & 25.177 & 0.347 & 0.58 & 0.82 &  0.07 & -1.00 & ab \\ 
V24  & 1:16:25.30 & 33:26:22.9 & 0.697 & 24.857 & 25.283 & 0.458 & 0.86 & 1.21 & -0.09 & -2.35 & ab \\ 
V25  & 1:16:26.21 & 33:26:54.2 & 0.519 & 25.021 & 25.426 & 0.464 & 1.17 & 1.65 & -0.02 & -1.63 & ab \\ 
V26  & 1:16:25.52 & 33:26:26.2 & 0.696 & 24.862 & 25.228 & 0.376 & 0.47 & 0.66 & -0.02 & -1.83 & ab \\ 
V27  & 1:16:25.59 & 33:26:25.5 & 0.509 & 24.868 & 25.216 & 0.381 & 0.89 & 1.25 &  0.04 & -1.19 & ab \\ 
V28  & 1:16:26.14 & 33:26:36.9 & 0.549 & 24.796 & 25.239 & 0.458 & 0.58 & 0.81 &  0.07 & -1.07 & ab \\ 
V29  & 1:16:27.08 & 33:26:43.4 & 0.385 & 24.914 & 25.163 & 0.261 & 0.47 & 0.67 & \nodata & \nodata & c \\ 
V30  & 1:16:25.41 & 33:25:44.1 & 0.545 & 24.861 & 25.308 & 0.462 & 0.55 & 0.78 &  0.07 & -1.00 & ab \\ 
V31  & 1:16:26.36 & 33:26:06.9 & 0.505 & 24.915 & 25.270 & 0.369 & 0.51 & 0.73 &  0.11 & -0.65 & ab \\ 
V32  & 1:16:26.25 & 33:25:55.3 & 0.651 & 24.824 & 25.248 & 0.429 & 0.36 & 0.51 &  0.03 & -1.43 & ab \\ 
V33  & 1:16:27.66 & 33:26:44.2 & 0.517 & 24.872 & 25.316 & 0.463 & 0.63 & 0.88 &  0.08 & -0.90 & ab \\ 
V34  & 1:16:27.53 & 33:26:32.0 & 0.449 & 24.744 & 25.023 & 0.334 & 1.13 & 1.60 &  0.05 & -1.02 & ab \\ 
V35  & 1:16:27.52 & 33:26:20.7 & 0.483 & 24.794 & 25.068 & 0.324 & 1.07 & 1.52 &  0.03 & -1.22 & ab \\ 
V36  & 1:16:26.98 & 33:25:57.5 & 0.538 & 25.085 & 25.392 & 0.329 & 0.66 & 0.93 &  0.06 & -1.09 & ab \\ 
V37  & 1:16:27.07 & 33:25:44.7 & 0.751 & 25.017 & 25.271 & 0.266 & 0.51 & 0.72 & -0.06 & -2.18 & ab \\ 
V38  & 1:16:28.21 & 33:26:15.7 & 0.695 & 24.897 & 25.301 & 0.441 & 0.94 & 1.33 & -0.10 & -2.45 & ab \\ 
V39  & 1:16:27.35 & 33:25:34.2 & 0.509 & 24.902 & 25.126 & 0.292 & 1.13 & 1.61 &  0.00 & -1.50 & ab \\ 
V40  & 1:16:29.17 & 33:26:37.2 & 0.665 & 24.895 & 25.235 & 0.364 & 0.74 & 1.04 & -0.04 & -2.01 & ab \\ 
V41  & 1:16:27.73 & 33:25:42.4 & 0.566 & 24.897 & 25.253 & 0.405 & 1.07 & 1.51 & -0.04 & -1.83 & ab \\ 
V42  & 1:16:28.68 & 33:26:12.0 & 0.341 & 24.780 & 25.214 & 0.454 & 0.59 & 0.83 & \nodata & \nodata & c \\ 
V43  & 1:16:28.77 & 33:26:06.3 & 0.490 & 24.931 & 25.280 & 0.379 & 0.85 & 1.19 &  0.07 & -0.99 & ab \\ 
V44  & 1:16:25.63 & 33:26:23.0 & 0.572 & 24.829 & 25.271 & 0.452 & 0.47 & 0.66 &  0.07 & -1.08 & ab \\ 
V45  & 1:16:27.89 & 33:26:12.1 & 0.797 & 24.842 & 25.310 & 0.473 & 0.34 & 0.48 & -0.05 & -2.18 & ab \\ 
V46  & 1:16:25.41 & 33:25:37.8 & 0.620 & 24.833 & 25.203 & 0.414 & 0.95 & 1.34 & -0.05 & -2.02 & ab \\ 
V47  & 1:16:26.77 & 33:25:31.4 & 0.763 & 24.807 & 25.140 & 0.379 & 1.01 & 1.43 & -0.09 & -2.32 & ab \\ 
V48  & 1:16:22.39 & 33:25:44.2 & 0.736 & 24.873 & 25.335 & 0.483 & 0.65 & 0.92 & -0.07 & -2.28 & ab \\ 
V49  & 1:16:24.02 & 33:25:30.0 & 0.544 & 24.919 & 25.314 & 0.409 & 0.54 & 0.77 &  0.08 & -0.98 & ab \\ 
V50  & 1:16:23.68 & 33:25:29.1 & 0.575 & 24.708 & 25.112 & 0.420 & 0.64 & 0.90 &  0.04 & -1.32 & ab \\ 
V51  & 1:16:27.10 & 33:25:10.7 & 0.673 & 24.706 & 25.093 & 0.422 & 0.91 & 1.29 & -0.08 & -2.29 & ab \\ 
V52  & 1:16:24.66 & 33:25:20.8 & 0.472 & 24.810 & 25.181 & 0.399 & 0.75 & 1.07 &  0.10 & -0.71 & ab \\ 
V53  & 1:16:22.98 & 33:25:26.2 & 0.316 & 24.861 & 25.246 & 0.398 & 0.47 & 0.66 & \nodata & \nodata & c \\ 
V54  & 1:16:25.18 & 33:25:15.7 & 0.454 & 24.916 & 25.263 & 0.380 & 0.88 & 1.24 &  0.09 & -0.73 & ab \\ 
V55  & 1:16:24.02 & 33:25:19.7 & 0.595 & 24.833 & 25.151 & 0.355 & 0.85 & 1.20 & -0.02 & -1.73 & ab \\ 
V56  & 1:16:24.77 & 33:25:12.5 & 0.604 & 24.921 & 25.333 & 0.446 & 0.90 & 1.27 & -0.03 & -1.86 & ab \\ 
V57  & 1:16:26.63 & 33:25:01.4 & 0.496 & 24.944 & 25.238 & 0.351 & 1.15 & 1.63 &  0.01 & -1.43 & ab \\ 
V58  & 1:16:25.83 & 33:25:01.4 & 0.282 & 24.839 & 25.232 & 0.403 & 0.41 & 0.58 & \nodata & \nodata & c \\ 
V59  & 1:16:23.98 & 33:25:06.1 & 0.557 & 24.830 & 25.178 & 0.355 & 0.42 & 0.59 &  0.09 & -0.91 & ab \\ 
V60  & 1:16:25.03 & 33:24:58.4 & 0.560 & 24.912 & 25.182 & 0.322 & 1.08 & 1.53 & -0.03 & -1.81 & ab \\ 
V61  & 1:16:23.33 & 33:25:03.4 & 0.442 & 24.871 & 25.200 & 0.372 & 0.92 & 1.30 &  0.10 & -0.68 & ab \\ 
V62  & 1:16:26.12 & 33:24:44.7 & 0.466 & 24.886 & 25.143 & 0.310 & 1.11 & 1.57 &  0.04 & -1.14 & ab \\ 
V63  & 1:16:24.53 & 33:24:51.5 & 0.502 & 24.797 & 25.038 & 0.289 & 1.05 & 1.48 &  0.02 & -1.33 & ab \\ 
V64  & 1:16:26.15 & 33:24:43.5 & 0.530 & 24.823 & 25.178 & 0.423 & 1.17 & 1.65 & -0.03 & -1.71 & ab \\ 
V65  & 1:16:26.06 & 33:24:41.3 & 0.686 & 24.907 & 25.244 & 0.384 & 0.95 & 1.35 & -0.10 & -2.41 & ab \\ 
V66  & 1:16:24.06 & 33:24:49.3 & 0.467 & 24.883 & 25.219 & 0.378 & 0.94 & 1.34 &  0.07 & -0.92 & ab \\ 
V67  & 1:16:24.77 & 33:24:45.2 & 0.466 & 24.900 & 25.231 & 0.386 & 1.06 & 1.50 &  0.05 & -1.07 & ab \\ 
V68  & 1:16:21.58 & 33:24:57.8 & 0.736 & 24.824 & 25.240 & 0.423 & 0.41 & 0.58 & -0.03 & -1.97 & ab \\ 
V69  & 1:16:22.26 & 33:24:52.9 & 0.430 & 24.737 & 25.128 & 0.404 & 0.48 & 0.67 & \nodata & \nodata & c \\ 
V70  & 1:16:22.28 & 33:24:50.2 & 0.707 & 25.011 & 25.474 & 0.475 & 0.52 & 0.73 & -0.03 & -1.96 & ab \\ 
V71  & 1:16:25.65 & 33:24:32.5 & 0.474 & 24.958 & 25.358 & 0.421 & 0.70 & 0.99 &  0.11 & -0.66 & ab \\ 
V72  & 1:16:20.70 & 33:24:50.3 & 0.469 & 24.841 & 25.190 & 0.390 & 0.99 & 1.40 &  0.06 & -1.00 & ab \\ 
V73  & 1:16:23.43 & 33:24:37.9 & 0.698 & 24.761 & 25.128 & 0.432 & 1.12 & 1.58 & -0.08 & -2.19 & ab \\ 
\enddata
\end{deluxetable}

\begin{deluxetable}{cccccc} 
\tablewidth{0pt} 
\footnotesize 
\tablecaption{Photometry of the Variable Stars ($B$)\label{tbl-3}}
\tablehead{
\colhead{} & \multicolumn{2}{c}{V01} & & \multicolumn{2}{c}{V02} \\ 
\cline{2-3} \cline{5-6} \\ 
\colhead{HJD-2450000} & \colhead{$B$} & \colhead{$\sigma_{B}$} & & 
\colhead{$B$} & \colhead{$\sigma_{B}$} 
          }
\startdata 
690.268 &  25.118  &  0.156 &&  \nodata  &  \nodata \\
690.319 &  25.254  &  0.117 &&  24.594  &  0.107 \\
690.336 &  25.285  &  0.164 &&   0.000  &  0.000 \\
690.386 &  25.288  &  0.139 &&  24.867  &  0.108 \\
690.403 &  \nodata  &  \nodata &&  25.020  &  0.108 \\
690.454 &  24.921  &  0.080 &&  25.183  &  0.119 \\
690.471 &  24.930  &  0.124 &&  25.171  &  0.108 \\
695.226 &  25.348  &  0.131 &&  24.916  &  0.103 \\
695.242 &  25.519  &  0.154 &&  24.835  &  0.087 \\
695.293 &  25.156  &  0.061 &&  \nodata  &  \nodata \\
695.310 &  25.090  &  0.106 &&  25.007  &  0.182 \\
695.360 &  24.828  &  0.127 &&  25.198  &  0.214 \\
695.377 &  \nodata  &  \nodata &&  25.284  &  0.137 \\
695.385 &  24.880  &  0.121 &&  25.325  &  0.115 \\
695.444 &  24.963  &  0.144 &&  25.267  &  0.140 \\
\enddata 
\tablecomments{The complete version of this table is in the electronic 
edition of the Journal.  The printed edition contains only a sample.}
\end{deluxetable}

\begin{deluxetable}{cccccc} 
\tablewidth{0pt} 
\footnotesize 
\tablecaption{Photometry of the Variable Stars ($V$)\label{tbl-4}}
\tablehead{
\colhead{} & \multicolumn{2}{c}{V01} & & \multicolumn{2}{c}{V02} \\ 
\cline{2-3} \cline{5-6} \\ 
\colhead{HJD-2450000} & \colhead{$V$} & \colhead{$\sigma_{V}$} & & 
\colhead{$V$} & \colhead{$\sigma_{V}$} 
          }
\startdata 
690.185 &  24.712  &  0.164 &&  25.147  &  0.171 \\
690.201 &  24.779  &  0.061 &&  25.202  &  0.106 \\
690.251 &  \nodata  &  \nodata &&  25.085  &  0.090 \\
695.092 &  \nodata  &  \nodata &&  25.109  &  0.118 \\
695.108 &  \nodata  &  \nodata &&  25.019  &  0.119 \\
695.158 &  24.894  &  0.097 &&  25.194  &  0.097 \\
695.174 &  24.956  &  0.091 &&  24.983  &  0.074 \\
\enddata 
\tablecomments{The complete version of this table is in the electronic 
edition of the Journal.  The printed edition contains only a sample.}
\end{deluxetable}

\begin{deluxetable}{cccccccccrc} 
\tablewidth{0pt} 
\tabletypesize{\scriptsize} 
\tablecaption{Frequency of Anomalous Cepheids in Dwarf Spheroidal Galaxies
\label{tbl-5}}
\tablehead{
\colhead{Galaxy} & \colhead{$\langle {\rm [Fe/H]} \rangle$} & 
\colhead{$M_V$} & \colhead{PA} & \colhead{$1-b/a$} & 
\colhead{$r_{\rm exp}$} & \colhead{Surveyed L} & 
\colhead{$N_{\rm AC}$} & \colhead{$S$} & \colhead{$\log\,S$} & 
\colhead{References}
\\
 & & & (degrees) & & (arcmin) & (\%) & & & 
          } 
\startdata 
Ursa Minor & --2.2 &  --8.9 &  53 & 0.55 &  5.8 &  90 &  6 & 2.3  & $0.36^{+0.15}_{-0.23}$  & 1,2,3,14   \\
Carina     & --2.0 &  --8.9 &  65 & 0.33 &  5.5 & 100 & 15 & 4.8  & $0.68^{+0.10}_{-0.13}$  & 1,2,15,18  \\
Draco      & --2.0 &  --8.9 &  82 & 0.29 &  4.5 &  55 &  5 & 3.8  & $0.58^{+0.16}_{-0.26}$  & 1,2,13     \\
Leo II     & --1.9 & --10.2 &  12 & 0.13 &  1.5 & 100 &  4 & 0.40 & $-0.40^{+0.18}_{-0.30}$ & 1,2,12     \\ 
Sculptor   & --1.8 & --10.7 &  99 & 0.34 &  6.5 &  55 &  3 & 0.35 & $-0.46^{+0.20}_{-0.36}$ & 1,2,3,16   \\ 
Sextans    & --1.7 & --10.0 &  56 & 0.40 &  9.1 &  75 &  6 & 1.0  & $0.00^{+0.18}_{-0.30}$  & 1,2,3      \\
Leo I      & --1.7 & --11.7 &  79 & 0.30 &  1.6 & 100 & 12 & 0.30 & $-0.52^{+0.11}_{-0.16}$ & 1,2,3,5,11 \\
Fornax     & --1.3 & --13.7 &  48 & 0.30 & 10.5 &  34 & 18 & 0.22 & $-0.66^{+0.09}_{-0.11}$ & 1,3,8,9,10 \\
 & & & & & & & & & & \\
And VI     & --1.6 & --11.3 & 160 & 0.23 &  1.4 &  21 &  6 & 1.1  & $0.04^{+0.14}_{-0.24}$  & 4,6,7,8    \\
And II     & --1.5 & --11.7 & 155 & 0.30 &  1.6 &  18 &  1 & 0.14 & $-0.85^{+0.31}_{-inf}$  & 3,8,17     \\
\enddata
\tablecomments{PA is the position angle measured from the North to the East.  
References:  (1) Mateo (1998); (2) Mateo et al.\ (1995); (3) Caldwell et al.\ 
(1992); (4) Caldwell (1999); (5) Gallart et al.\ (1999); (6) Armandroff et al.\ 
(1999); (7) Pritzl et al.\ (2002); (8) This paper; (9) Bersier \& Wood (2002); 
(10) Light et al.\ (1986); (11) Hodge \& Wright (1978); (12) Siegel \& 
Majewski (2000); (13) Baade \& Swope (1961); (14) Nemec et al.\ (1988); 
(15) Saha et al.\ (1986); (16) Kaluzny et al.\ (1995); (17) Da~Costa et al.\ 
(2000); (18) Dall'Ora et al.\ (2003)} 
\end{deluxetable}

\end{document}